\documentclass[aps,prb,showpacs,floatfix,superscriptaddress]{revtex4}
\usepackage{graphicx}

\begin{document}
\flushbottom

\title{Tunneling spectroscopy of the superconducting state of URu$_2$Si$_{2}$}
\author{A. Maldonado}
\affiliation{Laboratorio de Bajas Temperaturas, Departamento de
F\'isica de la Materia Condensada \\ Instituto de Ciencia de
Materiales Nicol\'as Cabrera, Facultad de Ciencias \\ Universidad
Aut\'onoma de Madrid, 28049 Madrid, Spain}
\author{I. Guillamon}
\affiliation{Laboratorio de Bajas Temperaturas, Departamento de
F\'isica de la Materia Condensada \\ Instituto de Ciencia de
Materiales Nicol\'as Cabrera, Facultad de Ciencias \\ Universidad
Aut\'onoma de Madrid, 28049 Madrid, Spain}
\author{J.G. Rodrigo}
\affiliation{Laboratorio de Bajas Temperaturas, Departamento de
F\'isica de la Materia Condensada \\ Instituto de Ciencia de
Materiales Nicol\'as Cabrera, Facultad de Ciencias \\ Universidad
Aut\'onoma de Madrid, 28049 Madrid, Spain}
\author{H. Suderow}
\email[Corresponding author: ]{hermann.suderow@uam.es}
\affiliation{Laboratorio de Bajas Temperaturas, Departamento de
F\'isica de la Materia Condensada \\ Instituto de Ciencia de
Materiales Nicol\'as Cabrera, Facultad de Ciencias \\ Universidad
Aut\'onoma de Madrid, 28049 Madrid, Spain}
\author{S. Vieira}
\affiliation{Laboratorio de Bajas Temperaturas, Departamento de
F\'isica de la Materia Condensada \\ Instituto de Ciencia de
Materiales Nicol\'as Cabrera, Facultad de Ciencias \\ Universidad
Aut\'onoma de Madrid, 28049 Madrid, Spain}
\author{D. Aoki}
\affiliation{INAC, SPSMS, CEA Grenoble, 38054 Grenoble, France}
\author{J. Flouquet}
\affiliation{INAC, SPSMS, CEA Grenoble, 38054 Grenoble, France}

\begin{abstract}
We present measurements of the superconducting gap of URu$_2$Si$_2$
made with scanning tunneling microscopy (STM) using a
superconducting tip of Al. We find tunneling conductance curves with
a finite value at the Fermi level. The density of states is V shaped
at low energies, and the quasiparticle peaks are located at values
close to the expected superconducting gap from weak coupling BCS
theory. Our results point to rather opened gap structures and gap
nodes on the Fermi surface.
\end{abstract}

 \date{\today}
\maketitle

\section{Introduction}

Superconductivity emerges out of heavy fermions in a number of
materials\cite{FlouquetRoad,Pfleiderer09}. Yet some of the basic
fundamental properties of heavy fermion superconductors remain
uncharacterized. Results from macroscopic measurements such as
specific heat or thermal conductivity imply that the superconducting
gap has zeros in some parts of the Fermi surface, forming the much
discussed line or point nodes characteristic of unconventional or
reduced symmetry superconductivity\cite{Mineev}. Unconventional
superconductivity is indeed likely to be favored within strongly
correlated heavy electrons, to avoid mutual electron repulsion in
the formation of Cooper pairs.

Recently, the application of scanning tunneling microscopy (STM)
technique to heavy fermions has brought the field a significant step
further\cite{Haule09,Schmidt10,Aynajian10,Hamidian11,Suderow04,Fischer07,Suderow08Pr,Ernst10}.
Attention has been turned to the so called hidden order state of
URu$_2$Si$_2$, with the synthesis of new generation of high quality
ultraclean samples in this compound \cite{Matsuda11}. The low
temperature hidden order (HO) phase of URu$_2$Si$_2$ is indeed
characterized by a low carrier density and huge entropy changes with
a microscopic ordering whose nature is not yet
determined\cite{Mydosh11}. If this phase was antiferromagnetically
ordered, the moment magnitude would be far too small (0.02 $\mu_B$)
to account for the large entropy changes. Thus, antiferromagnetism
(AF) is believed to be not an intrinsic phenomena\cite{Matsuda01}
but an extrinsic residual component due to the proximity of AF
induced at rather low pressure (0.5GPa) or uniaxial
stress\cite{Amitsuka07,Hassinger08,Bourdarot11}. More complex
multipole ordering have been proposed: octupole\cite{Kiss05},
hexadecapole\cite{Haule09,Kusunose11} and
dotriaecutapole\cite{Cricchio09} with even a nematic
character\cite{Ikeda_tobepublished}. Helical Pomeranchuk
order\cite{Varma06}, modulated spin liquid\cite{Pepin11}
hybridization wave\cite{Dubi11}, and rank-5 spin density wave\cite{Rau12} have been considered. Early hints
towards quadrupolar ordering have been more recently excluded using
detailed experiments\cite{Rodrigo97,Amitsuka10}. The evolution of a
gap opening when entering the low temperature HO phase at
T$_{HO}$=17.5 K has been followed in detail in atomically resolved
experiments\cite{Haule09,Schmidt10,Aynajian10,Hamidian11}.
Hybridization between heavy quasiparticles as viewed from scattering
in Th doped samples has been discussed in terms of interference
effects between multiple channel tunneling\cite{Ternes09,Yuan11}.

This peculiar ground state hosts a superconducting phase inside,
whose properties are ill-known, in spite of much work since its
discovery in 1985\cite{Palstra85}. Thus, the simple characterization
of the superconducting state remains of great
importance\cite{Amitsuka07,Hassinger08,Soltan11,Oppeneer10}. The
presence of nodes in the superconducting gap seems well established
from several macroscopic experiments, such as specific heat or
thermal conductivity\cite{Knetsch92,Brison94,Yano08,Kasahara09}.
Evidence for multiple gaps has been provided notably by the upper
superconducting critical field\cite{Dubi11,Brison95}. This is not
very surprising, as the band structure is complex with sheets
showing different mass renormalizations\cite{Jo07,Oppeneer10}. The
measurement of the tunneling spectroscopy of the superconducting
density of states has eluded until now all experimental attacks,
although early point contact experiments yielded some
insight\cite{Hasselbach92,Naidyuk96}. Here we provide successful
tunneling spectroscopy results in the superconducting phase. We
determine values for the superconducting gap and its temperature
dependence, and find a density of states, which changes as a
function of the position, and has a finite value at the Fermi level.

\section{Experimental}

We use a STM device in a MX400 dilution refrigerator of Oxford
Instruments built and tested following previous
work\cite{Suderow11,Crespo06a}. We use superconducting tips of
Al\cite{Rodrigo04,Guillamon07}. These tips allow investigating, at
the same time, S-URu$_2$Si$_2$ and N-URu$_2$Si$_2$ tunneling curves
by measuring, respectively, at zero field and with a small magnetic
field above the critical field of the Al tip (of order of 0.04 T,
see Ref.\cite{Guillamon07}). The device features a positioning
system which allows to change the scanning window of 4 $\mu$m$^2$
in-situ, and to bring the tip to a sample of the same material for
cleaning and preparation. We measure the tunneling current $I$
across the junction while applying a bias voltage ramp $V$. The
tunneling conductance curves $dI/dV$ versus bias voltage are
obtained by numerical differentiation and are usually normalized to
the conductance obtained at a bias voltage a few times above the
voltage where superconducting features appear. Superconducting tips
of Al are prepared and cleaned in-situ on an Al pad by mechanical
annealing from repeated indentation. The scheme of the experiment
is presented in Fig.1a. The tip
is cleaned through an initial repeated indentation process, i.e. by inserting the tip into the sample and then making indentation and retraction repeatedly with an amplitude of some hundreds of nm using the z-piezo. When the tip is pulled out, after some thousands of repeated indentations, we observe clean conductance versus displacement curves which identify few atom Al contacts (see
Fig.1b)\cite{Rodrigo04,Guillamon07}. The tip is totally pulled out of the Al sample, and we observe superconducting S-S tunneling curves in the current vs bias voltage curves. In clean s-wave BCS S-S tunnel junctions, simplest s-wave BCS theory gives divergent quasiparticle peaks in the conductance. The width of the actually measured quasiparticle peaks can thus be associated to the voltage jitter of the experiment \cite{Suderow11,Crespo06a,Rodrigo04,Guillamon07,Maldonado11}, for which we find here 26$\mu$eV (Fig. 1c, lower panel).
Once the tip has been thus characterized and the associated conductance verified to follow expected S-S features, it is moved to a clean Au sample. The N-S tunneling conductance gives then a good measurement of the lowest temperature attained in the microscope \cite{Maldonado10}. Simple BCS s-wave fits of the experimental conductance curve (Fig.1c, upper panel), give 150 mK. Finally, the tip is brought to the sample of URu$_2$Si$_2$ for measurement.

\begin{figure}[ht]
\includegraphics[width=12cm,keepaspectratio,clip]{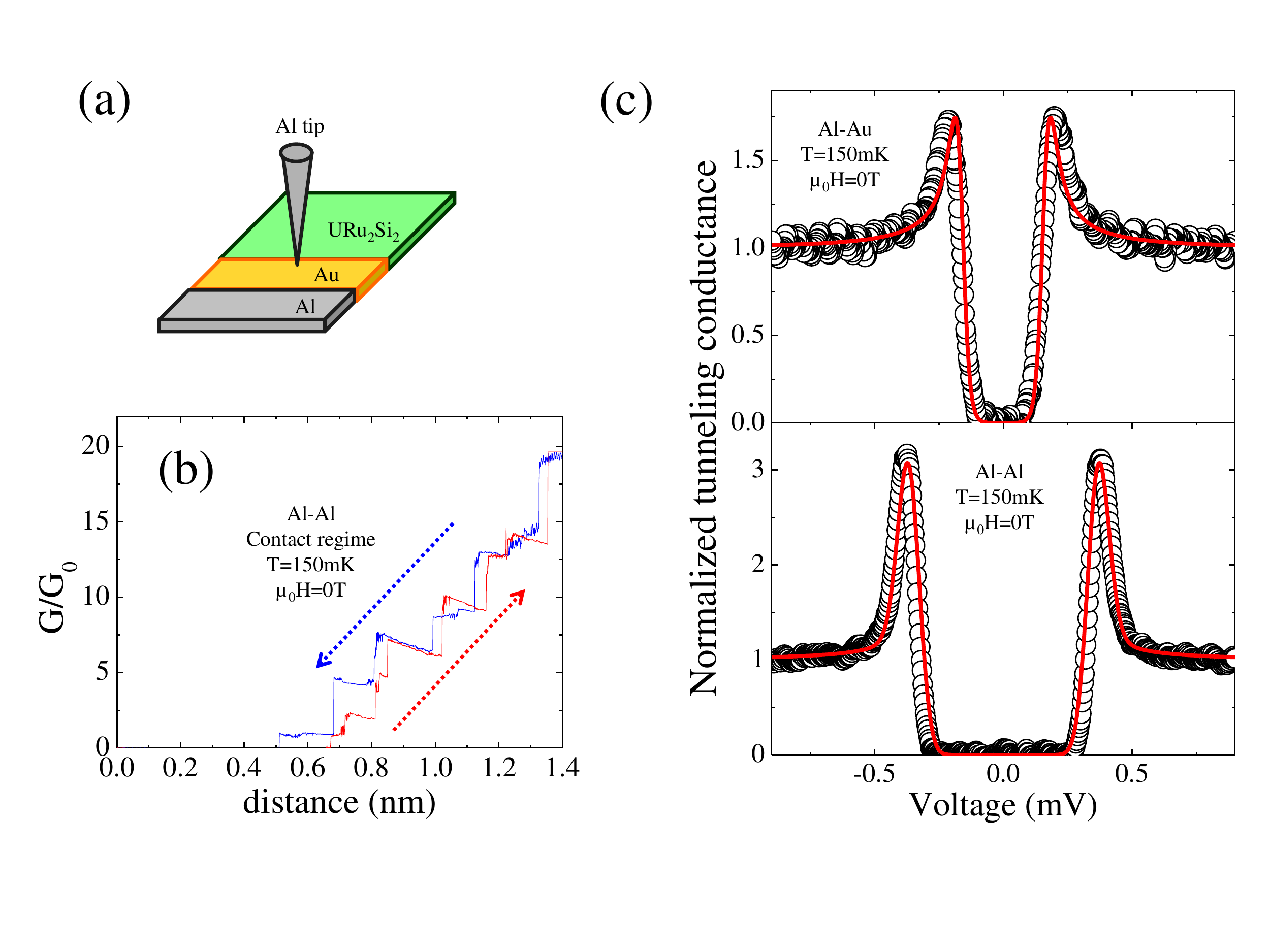}
\vskip -0cm \caption{(a) Scheme of the experiment. (b) Conductance
vs. displacement curves when retracting the tip after a continued
repeated indentation procedure made following
\protect\cite{Rodrigo04,Guillamon07}. Curves during indentation
(red) and tip retraction (blue) are highlighted by arrows. The
conductance shows the characteristic Al-Al single atom point contact
features, namely steps at single and few atom contacts at values
related to the quantum of conductance G$_0=2e^{2}/h$ with a negative
slope as plotted here (see \protect\cite{Rodrigo04} and references
therein). The work function in the tunneling limit (at G$<<$G$_0$,
i.e. well below 1$\mu S$), is of several eV. The whole curve shows
that the last atoms are of Al and that the tip can be used for clean
vacuum tunneling. (c) When the tip is separated from the sample,
clean Al-Al spectra are observed. S-S curves (bottom
panel) are taken when the Al tip is located above the Al sample. N-S
curves are obtained when the same tip is moved in-situ on top of an
Au sample (top panel). Red lines are fits to BCS using
$\Delta$=0.16meV, T=150mK and an energy resolution (bias voltage
jitter) of 26$\mu$eV.} \label{fig1}
\end{figure}

Single crystals of
URu$_2$Si$_2$ were grown by Czochralski method in a tetra-arc
furnace with argon gas atmosphere according to Ref.\cite{Matsuda11}.
We decided to break URu$_2$Si$_2$ samples along the basal plane of the tetragonal structure at room temperature ambient conditions immediately before mounting them on the STM and cooling down. This has the disadvantage that surface contamination cannot be totally avoided, but it also allows to use a set-up suitable for superconducting tips, where the procedures described above can be performed. Moreover, the surface can be optically inspected prior to cool down and samples with nice looking surfaces can be chosen. Many materials prepared ex-situ have been found to show clean tunneling features and bulk related spectroscopic curves with a high reproducibility. Succesful experiments have been made in ex-situ cleaved, as grown and even chemically etched surfaces\cite{Rodrigo97,Suderow08Pr,Maggio95,DeWilde,Fischer07}. Additionally, in our set-up we can macroscopically change at the lowest temperatures the scanning window without heating\cite{Suderow11}. This allows to search in-situ over the whole mm sized sample the scanning windows (of 4$\mu$m$^2$) with best tunneling conditions. In this way, we could find in URu$_2$Si$_2$ scanning windows with high work functions and a clean surface providing images which are independent of the value of the tunneling current. There, we often found irregular
surfaces, within which we could find small atomically flat areas. An example of the topography is shown in Fig.2. The atomic
lattice of the tetragonal basal plane can be resolved (see Fig.
2(b)), with height modulations whose period corresponds to the
expected basal plane lattice parameter. We observe high bias voltage Fano like features (inset of Fig.2a) due to interference between tunneling into light and heavy mass bands, characteristic of heavy fermions. These features appear above some mV and are comparable to those obtained previously \cite{Haule09,Schmidt10,Aynajian10,Hamidian11,Ernst10}. In this work we focus on
tunneling spectroscopy measurements at low bias voltage in order to
explore the superconducting phase of URu$_2$Si$_2$. The typical
junction conductance is of  0.4 $\mu$S and the bias voltage of about
1.25mV. The tunneling curves at zero field show clean S-S' tunneling
features, which go over into N-S' features when applying a magnetic
field of 0.1 T.

\begin{figure}[ht]
\includegraphics[width=12cm,keepaspectratio,clip]{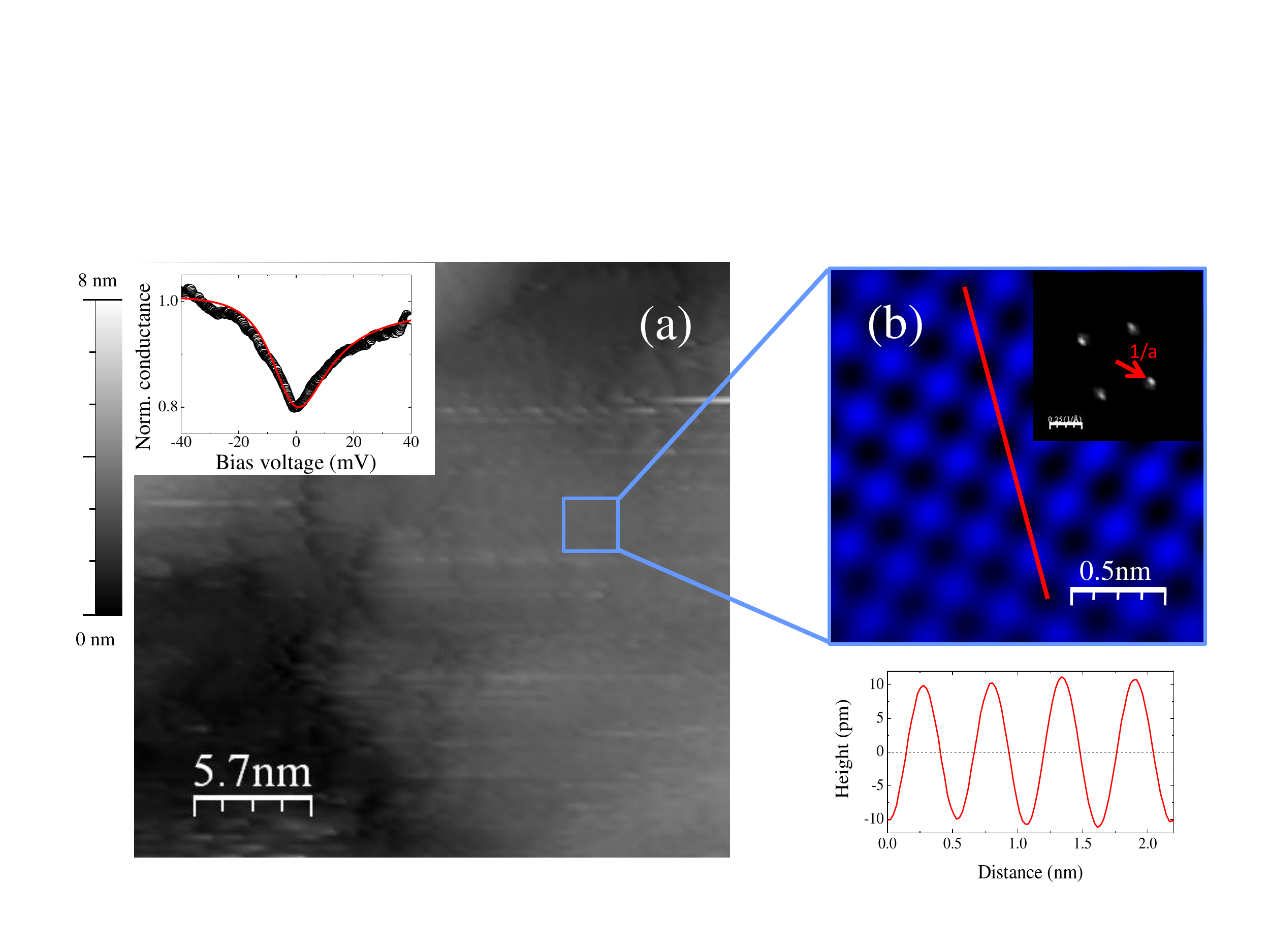}
\vskip -0cm \caption{(a) Example of the topography observed, showing
typical roughness of about 8nm. The image was taken at a
conductance of 0.4$\mu$S and a bias voltage of 1.25mV. In these
regions, small flat areas can be found, where atomic resolution can
be obtained.  Such regions show Fano like features in the high bias
voltage conductance, reassembling those obtained in previous work
over a terminated Ru surface (inset, the red line is a fit to Fano
function using an asymmetry parameter $q$=-0.2 and a width of
12meV)\protect\cite{Haule09,Schmidt10,Aynajian10,Hamidian11}. (b)
shows an atomic resolution image taken at a conductance of
0.27$\mu$S, and a bias voltage of 1.87 mV. It shows the basal plane
of the tetragonal structure, possibly of the Ru sublattice
\protect\cite{Haule09,Schmidt10,Aynajian10,Hamidian11}, and has been
Fourier filtered to reveal salient features at the reciprocal
lattice wavevector of 0.03 nm$^{-1}$. From the line scan in real
space shown in the bottom panel, we obtain a lattice parameter of
0.3 nm, possibly corresponding to the Ru interatomic distance.} \label{fig2}
\end{figure}

\section{Results and discussion.}

In Fig.3 we show some characteristic tunneling spectroscopy curves
obtained at different positions at zero magnetic field
(S-URu$_2$Si$_2$). The superconducting features of URu$_2$Si$_2$ are
clearly resolved in them. These features are homogeneous
over small flat areas, but, in different positions, we find slight
differences, as presented in Fig.3. They appear to be related to the
particular area where they were taken, and they change at scales of
about 10 nm. In the S-URu$_2$Si$_2$ normalized tunneling conductance
curves (Fig.3(a), bottom panels), we observe zero conductance at the
Fermi level, resulting from the zero density of states of Al, which
increases steeply above 0.2 mV, showing a pronounced shoulder. S-S'
conductance curves at very low temperatures between two s-wave BCS
superconductors have a zero conductance up to the sum of both gaps,
where a steep peak is found (Fig.1b and Refs.\cite{Wolf,Suderow02,Rodrigo04b}). The
presence of a marked shoulder in our experiments shows that the
density of states of URu$_2$Si$_2$ is different from a conventional
single gap s-wave superconductor.

The tunneling conductance between a superconducting tip and a sample
is, in most simple single particle models, given by $I(V) \propto
\int dE [f_T(E-eV)-f_S(E)] N_T(E-eV) N_S(E)$, where $N_T(E)$ and
$N_S(E)$ are the respective densities of states of tip and sample,
and $f_{T,S}$ the respective Fermi occupation functions. Using
previously determined $N_T(E)$ from the curves obtained when the tip
is on a normal Au sample (Fig.1(c), upper panel)
\cite{Rodrigo04,Guillamon07}, we can obtain $N_S(E)$ by
de-convolution from the integral, getting the curves shown in
Fig.3(b). The de-convoluted curve is of course smeared and does not
show noise fluctuations of the experimental
conductance\cite{Guillamon08}. Note the peculiar low energy
behavior, with a V-shaped form at low energies and well developed
quasiparticle peaks. This V- shape actually produces the shoulder
observed in the tunneling conductance, and shows a continuos
increase of the density of states from zero energy, as expected in a
superconductor with nodes in the gap function. The low but finite
zero energy density of states and its changes as a function of the
position can be related to band or orientation dependent tunneling
into zero gap regions. The topographic features of the surface can
also influence the results. More detailed measurements in large,
atomically resolved areas, will be helpful to explore this further.

Application of 0.1 T drives the tip into the normal state and reveals a tunneling
density of states characterized by a low
but finite value close to zero energy and wide quasiparticle peaks,
pointing towards a sizable distribution of values of the
superconducting gap (see Fig.4). Previous reported values of the
superconducting gap from tunneling spectroscopy give gap sizes
several times $\Delta_0$=1.73k$_B$T$_c$
\cite{Hasselbach92,Naidyuk96}. Here, instead, we observe
quasiparticle peaks located roughly at 0.30 meV, which is 1.36
$\Delta_0$. All this is in good agreement with the reduced jump of
the specific heat $\frac{\Delta C}{\gamma T_{c}}\sim 1$ derived by
entropy conservation\cite{Matsuda11}.

\begin{figure}[ht]
\includegraphics[width=12cm,keepaspectratio,clip]{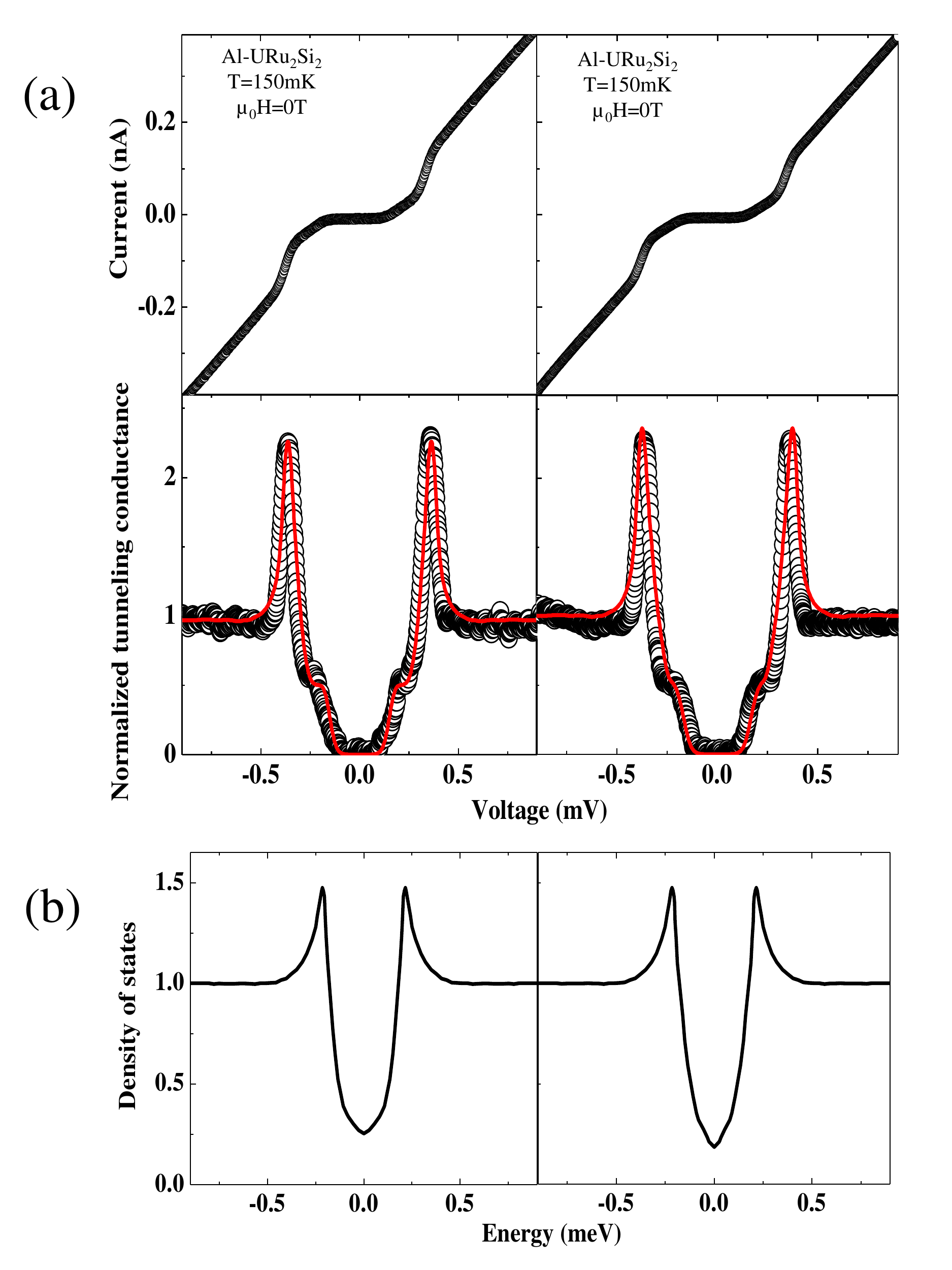}
\vskip -0cm \caption{(a) We show tunneling current vs bias voltage
(upper panels) and normalized tunneling conductance (lower panels)
curves obtained at 0.15 K and zero field, with the Al tip being
superconducting. Red lines in (a) are convolutions obtained using
the density of states of Al for the tip and the densities of states
shown in (b) for URu$_2$Si$_2$. Curves are obtained at two different
positions of the surface.} \label{fig3}
\end{figure}

The temperature dependence of N-URu$_2$Si$_2$ curves is shown in
Fig.5. Superconducting features of URu$_2$Si$_2$ disappear around
1.5 K. We can plot the temperature dependence of the density of
states at the Fermi level, and of the energy for which the
derivative of the density of states is maximum. Within experimental
uncertainty, the latter follows roughly simple BCS theory. In the
former, we do not find an appreciable temperature dependence within
the uncertainty of our method. Accuracy in its determination, and
the scatter found over different positions does not allow to
distinguish the small difference expected between the temperature
dependence of the gap value, $\Delta$, in a nodal superconductor and
in the most simple isotropic s-wave BCS superconductor\cite{Won94}.

\begin{figure}[ht]
\includegraphics[width=12cm,keepaspectratio,clip]{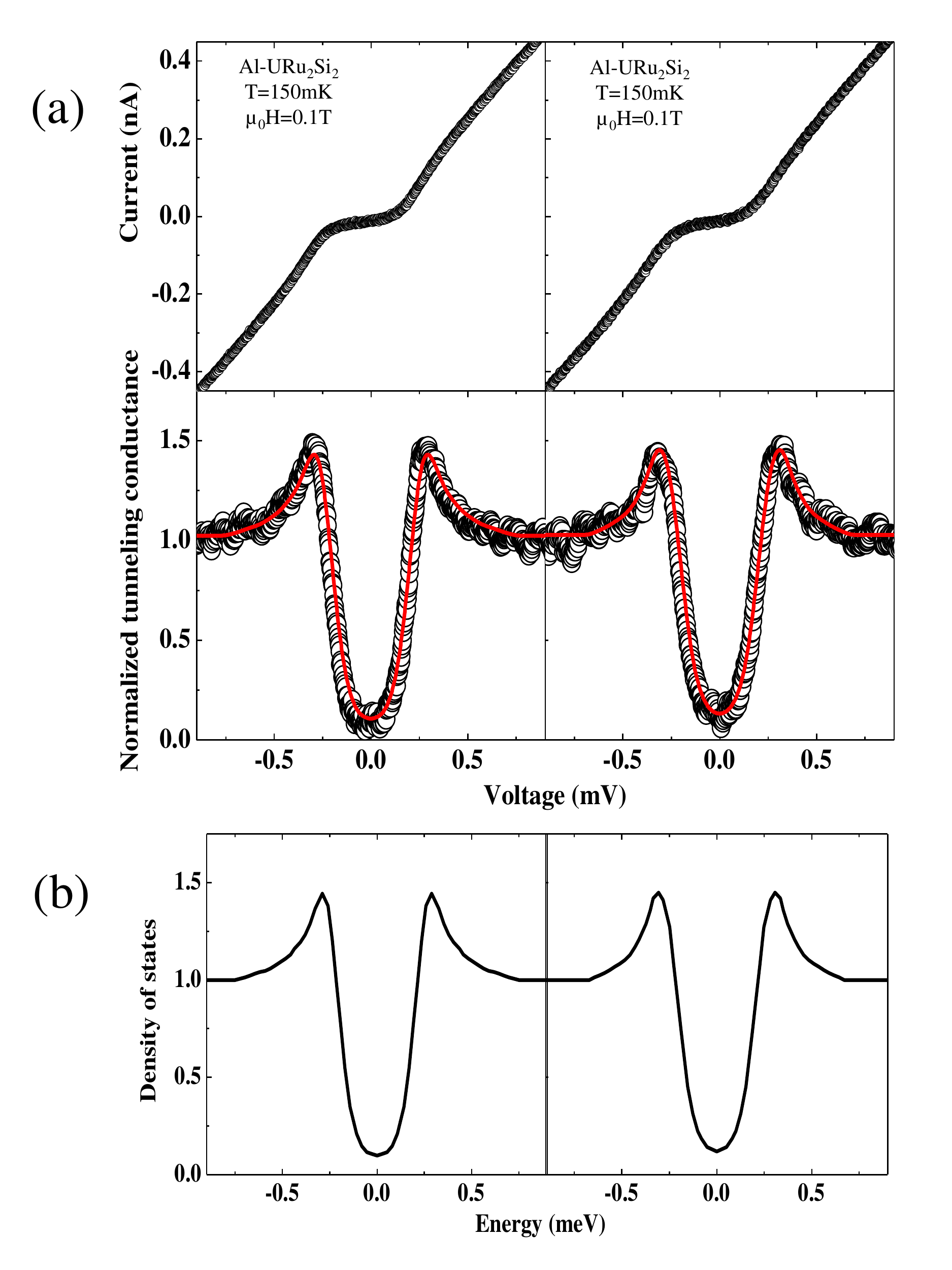}
\caption{(a) Tunneling current vs bias voltage (upper panels) and
normalized tunneling conductance curves (lower panels) obtained at
0.15 K by applying a magnetic field of 0.1 T, which drives the Al
tip to the normal state. Red lines in lower panels of (a) show the
tunneling conductance obtained using the densities of states shown
in (b). Curves are obtained at two different points on the surface.}
\label{fig4}
\end{figure}

Note that we find a small difference between the densities of states
generally found using superconducting tips at zero field than those
found using normal tips at 0.1 T. Remarkably, when the tip is
normal, the quasiparticle peaks are more rounded and located at 20\%
higher energies and an amount of states close to the Fermi level is
decreased by 10\%, leading, in some particular positions, to
apparently better developed BCS like curves. Interestingly, orbital
pair breaking effects by the magnetic field, such as the ones
produced by the presence of vortices close to the tip should lead to
a decrease in the size of the gap and an increase in the Fermi level
conductance, i.e. opposite as observed. Thus, either paramagnetic
effects appear in the density of states, or the nodes tend to close
and the gap opens in the presence of a magnetic field.

Note also that the method for obtaining the density of states
discussed in Figs.3-5 is the simplest approximation for tunneling
and assumes single particle tunneling. In a strongly correlated
heavy fermion, simultaneous tunneling into light and heavy masses,
as well as Fermi liquid effects can lead to substantial
modifications of the tunneling spectra, as mentioned previously and discussed in Refs.\cite{Yazdani97,Ternes09,Haule09,Schmidt10,Aynajian10,Hamidian11,Yuan11} 
in relation with measurement at energies above the
meV.
It is not yet clear how these features may influence superconducting
tunneling at smaller energy scales. A more detailed atomic size study simultaneously imaging high bias voltage and superconducting features would give more insight. Additionally, a theory of S-S'
tunneling where S' is a heavy fermion may help to establish a more direct relationship between the superconducting order parameter and the features observed here in the density of states,
namely, V-shape at low energies, finite value at zero energy, and
slight opening when applying a magnetic field.

\begin{figure}[ht]
\includegraphics[width=12cm,keepaspectratio,clip]{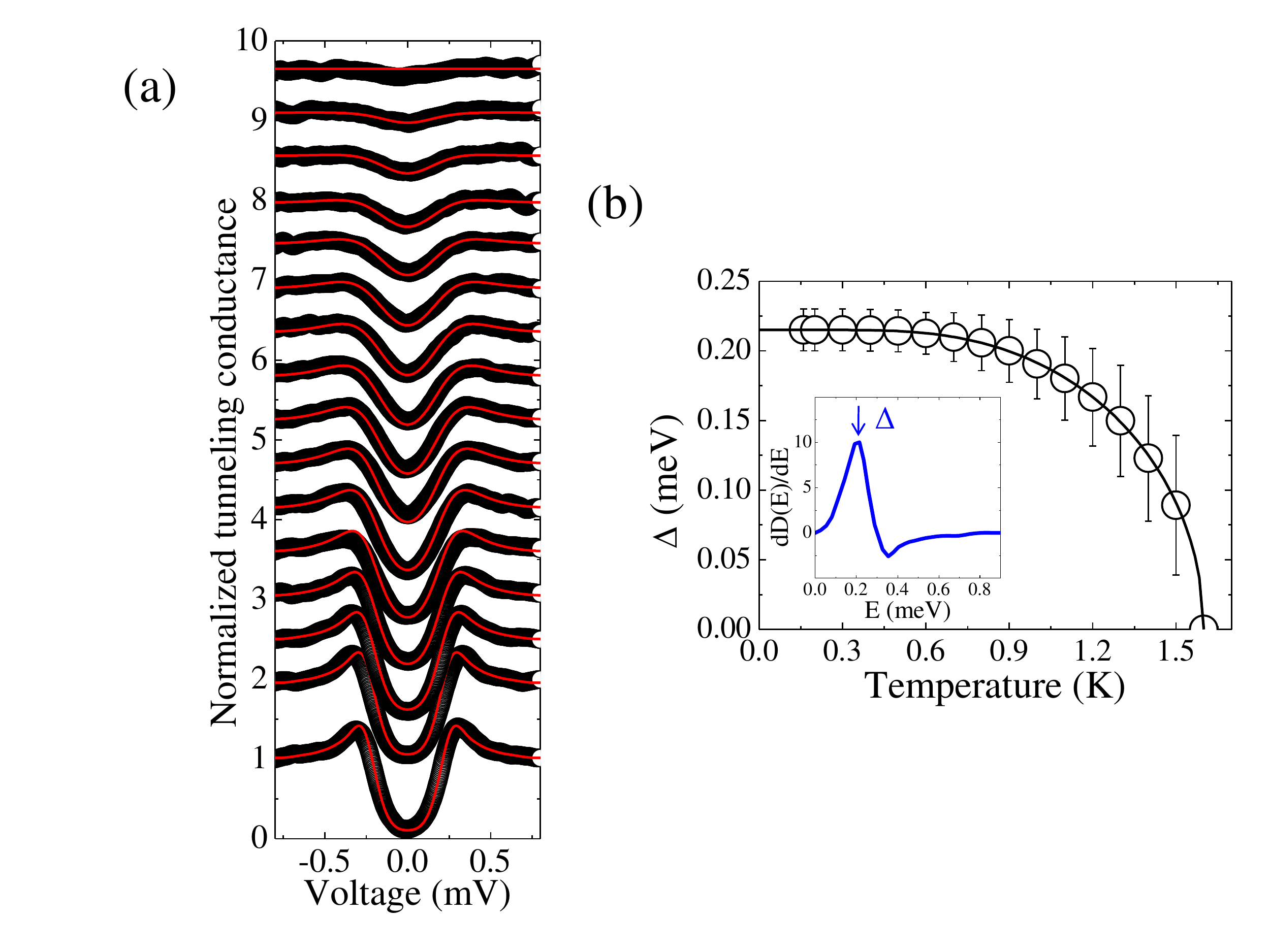}
\caption{The temperature dependence of the tunneling spectroscopy of
URu$_2$Si$_2$ at 0.1 T is shown in (a). The data are taken, from
bottom to top, at 0.15K and from 0.2 to 1.6K at steps of 0.1K. At
each temperature, we de-convolute the density of states, and make
its derivative as a function of energy, shown as a blue line for
0.15 K in the inset of (b). The inflexion point in the quasiparticle
peak of the density of states is shown by an arrow, and can be taken
as a good measure of the superconducting gap. Its temperature
dependence is shown in (b). The black line is a guide to the eye.
The error bars highlight the uncertainty in the determination of the
de-convolution, which increases together with temperature smearing.
Red lines in (a) are the conductance curves obtained using the
de-convoluted density of states \cite{Guillamon08}.} \label{fig5}
\end{figure}

Other features characteristic of S-S' junctions, such as the
Josephson effect, and the temperature dependence of the S-S'
conductance curves, require a detailed analysis and more
experiments. Note that the Josephson coupling energy at the
tunneling conductance used here is far below the thermal energy
k$_B$T, which brings the Josephson peak below the resolution of the
current measurement, even in a Al S-S junction\cite{Rodrigo04}.
Measurements at a lower tunneling conductance and as a function of
temperature are under way.

Finally, let us remark that often specific heat data show a
transition centered around 1.3-1.4 K, which is broad and strongly
featured\cite{Matsuda11,Kasahara07}. The specific heat increases
well above the midpoint of the transition, evidencing a reduced
entropy at temperatures close to or above 1.5 K. The origin of such
a featured transition has remained ill-understood. On the other
hand, when applying pressure, the resistivity retains
superconducting features above $\approx$ 0.7 GPa, where
antiferromagnetism appears\cite{Hassinger08}, but no bulk
superconductivity is observed in the specific heat above this
pressure. Our results show that the conductance remains gapped up to
1.5 K, and that the density of states changes in different points on
the surface. Remarkably, the residual term in the specific heat is
small, of the order of 10\%, agreeing with the observed low values
of the zero energy conductance observed here. Thus, the
superconducting behavior appears to be inhomogeneous, without a
significant pair breaking effect affecting the low energy density of
states. On the other hand, evidences for an anomalously low carrier
density have been provided, and related to the hidden order gap
opening\cite{Behnia05}. The low carrier density can make the
superconducting properties sensitive to structural distortion. A
complex order parameter, such as e.g. the suggestion of a "chiral"
state governed by the nematic HO phase\cite{Kasahara09}, could also
lead to inhomogeneous superconducting behavior.

\section{Conclusions.}

In summary, we have measured the tunneling spectroscopy in the
superconducting phase of URu$_2$Si$_2$ using very low temperature
scanning tunneling spectroscopy. Our results are a significant
instrumental step forward which will help to understand properties
of heavy fermion superconductors. Although we find rather irregular
surfaces, tunneling curves show opened gap structures, with values
of the order of the weak coupling gap (1.73k$_B$T$_c$) and a small
but finite density of states at the Fermi level. Finally, the finite
and V-shaped density of states observed close to the Fermi level, most clearly revealed when using of a superconducting tip at zero field, points
towards the presence of nodes along some directions of the Fermi
surface.

\section{Acknowledgments.}

The Laboratorio de Bajas Temperaturas is associated to the ICMM of
the CSIC. This work was supported by the Spanish MICINN and MEC
(Consolider Ingenio Molecular Nanoscience CSD2007-00010 program,
FIS2011-23488, ACI-2009-0905 and FPU grant), by the Comunidad de
Madrid through program Nanobiomagnet and by ERC (NewHeavyFermion),
and French ANR projects (CORMAT, SINUS, DELICE).


\end{document}